\documentclass[epjCONF]{svjour}

\usepackage{graphicx}
\usepackage[varg]{txfonts} 
\usepackage[latin1]{inputenc}

\session-title{40th Liege International Colloquium : Ageing low mass stars : from red giants to white dwarf.}

\begin{document}

\title{Non-radial, non-adiabatic solar-like oscillations in RGB and HB stars}
\author{M. Grosjean\inst{1} \inst{,3} \fnmsep \thanks{\email{grosjean@astro.ulg.ac.be}} \and M.A. Dupret\inst{1} \and K. Belkacem \inst{2} \and J. Montalban \inst{1} \and{A. Noels } \inst{1} \and R. Samadi \inst{2} }

\institute{Institut d'Astrophysique Géophysique et Oceanographie de l'Univertisté de Liège, Belgium \and LESIA, Observatoire de Paris-Meudon, France \and Doctorant, Boursier F.R.I.A.}

\abstract{
CoRoT and {\it Kepler} observations of red giants reveal rich spectra of non-radial solar-like oscillations allowing to probe their internal structure.
 We compare the theoretical spectrum of two red giants in the same region of the HR diagram but in different evolutionary phases.
We present here our first results on the inertia, lifetimes and amplitudes of the oscillations and discuss  the differences between the two stars.	
} 

\maketitle
\section{Introduction}\label{intro}
The two models considered in this study are a red giant branch (RGB) and a helium burning model (HB). Both were  computed with the code ATON \cite{RefAton}. Their mass is $1.4 M_\odot$ and their radius is $ 11.9 R_\odot$. MLT is used to described the convection with $\alpha_{MLT} = 1.9$ and the initial chemical composition is $X=0.7$ and $Z=0.02$.
To compute the mode lifetimes, we use the non-adiabatic pulsation code MAD \cite{RefMAD} with a non-local time-dependant treatment of the convection \cite{RefNL}. The amplitudes are computed using a stochastic excitation model \cite{RefStoc} and solar parameters to describe the turbulence in the envelope.

\section{Results}
\label{res}
\begin{figure}[h!]
\resizebox{0.9\columnwidth}{!}{%
 \includegraphics{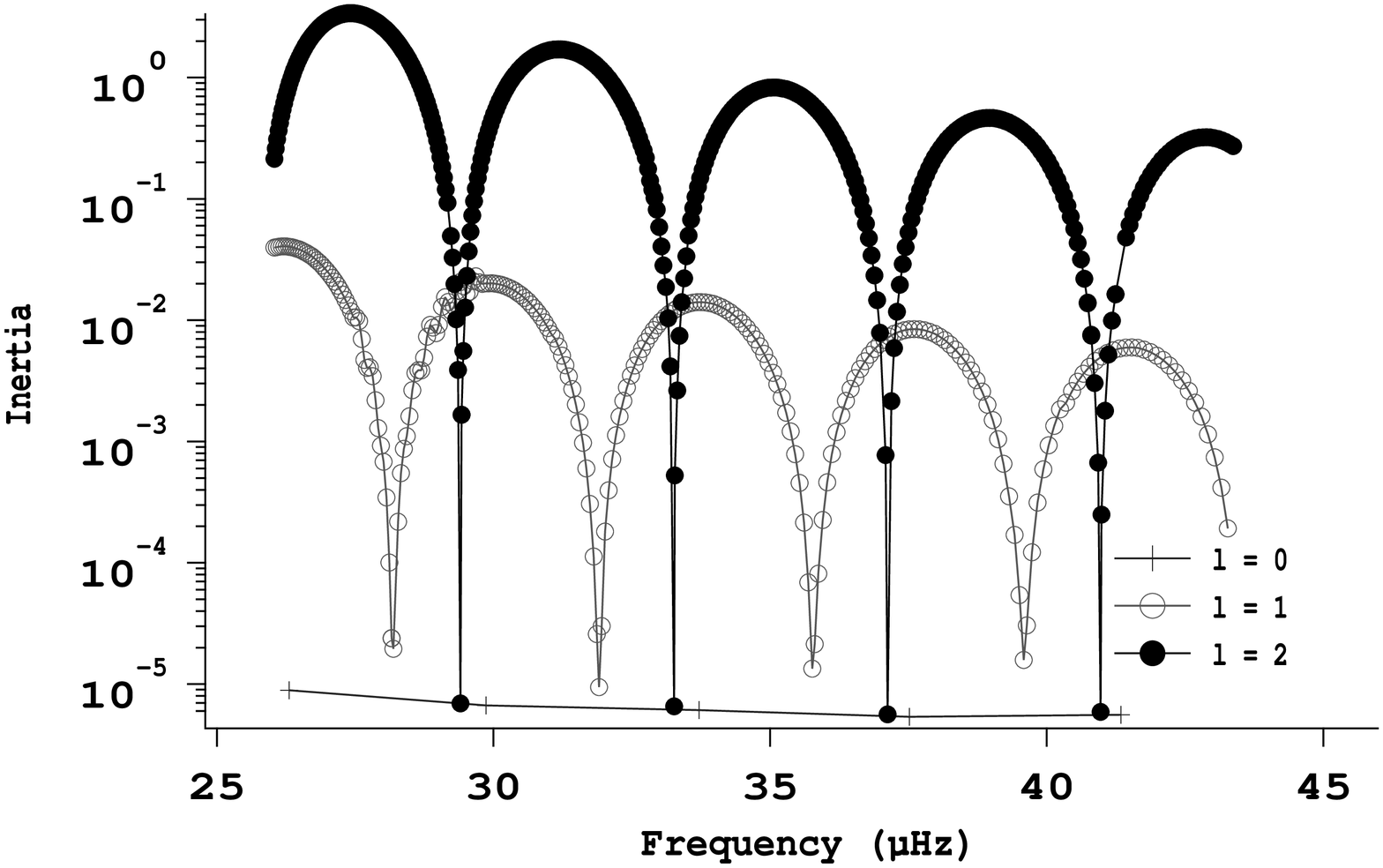}
 \includegraphics{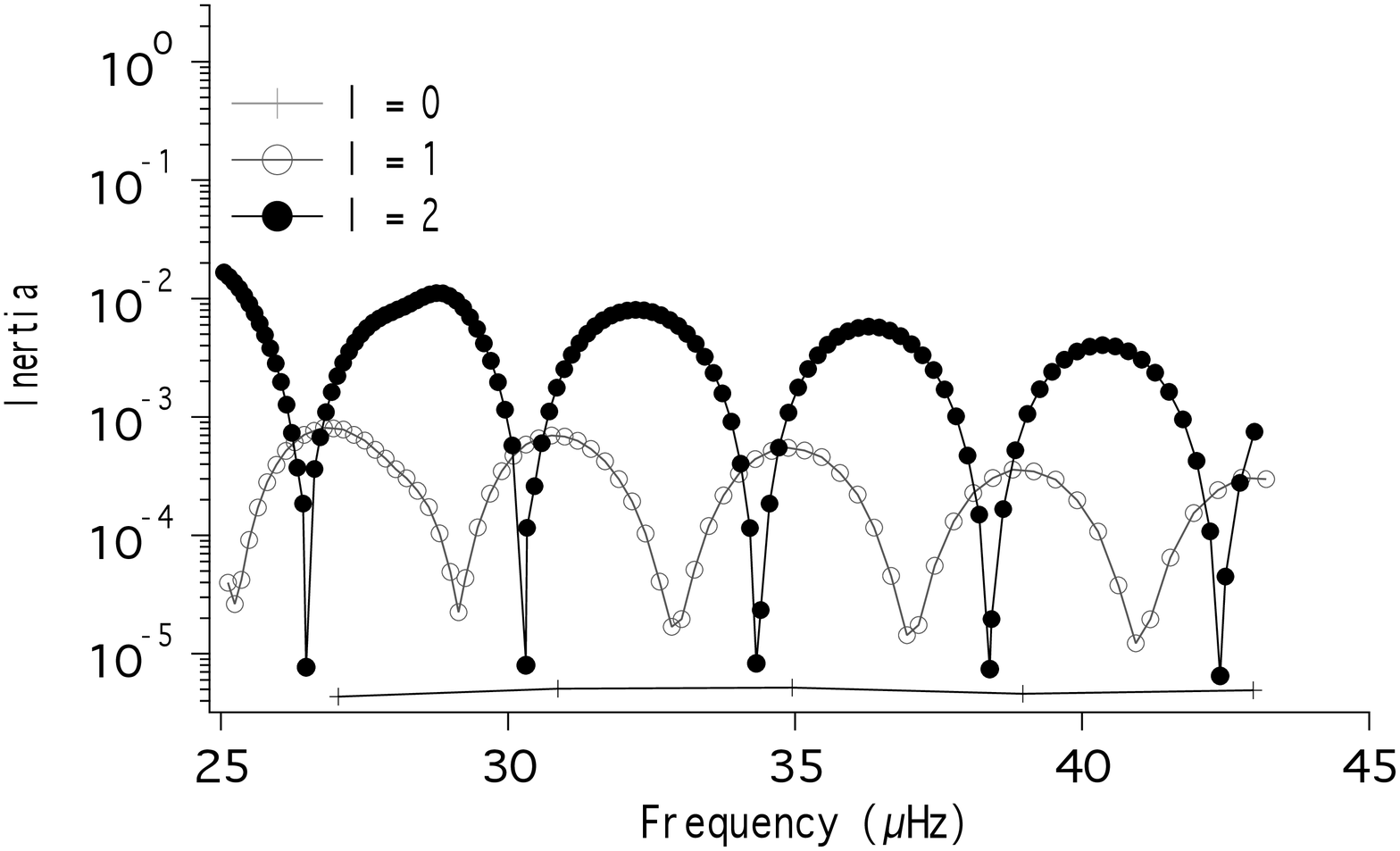} 
}
\caption{\textbf{Inertia} of l=0, 1 and 2 modes for the RGB model (left) and the HB model (right). 
Thanks to the inertia we can distinguish two type of non-radial modes.
 The modes with low value of inertia are trapped in the envelope (p-type behavior).
As in the adiabatic case the l=2 modes are better trapped in the envelope than the l=1 \cite{RefMontalban}.
We also note a higher density of modes and a better trapping in the RGB model (see also \cite{RefMontalban2}). 
 }
\label{fig:1}    
\end{figure}

\small{}

\begin{figure}[h!]
\resizebox{0.9\columnwidth}{!}{
\includegraphics{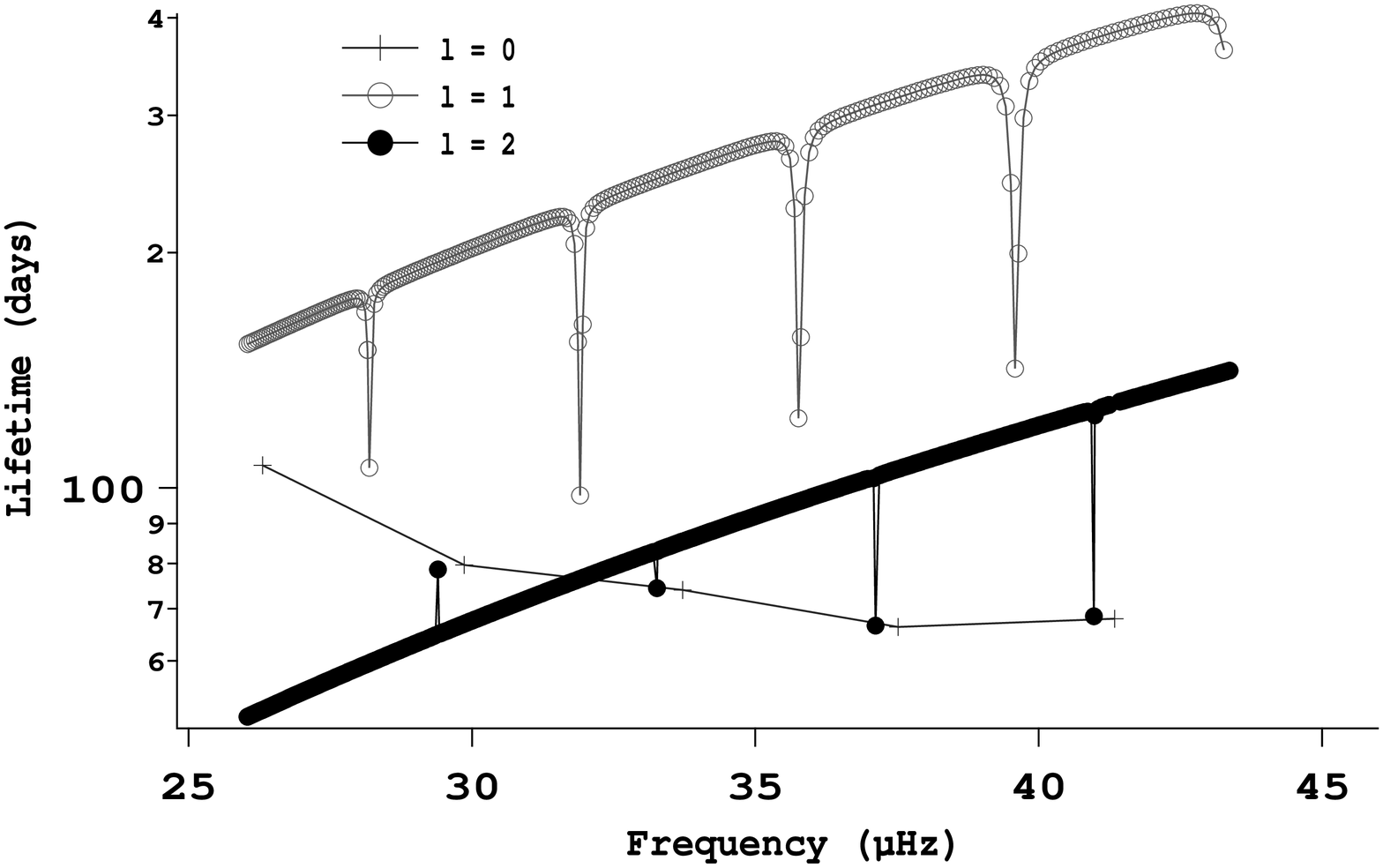} 
\includegraphics{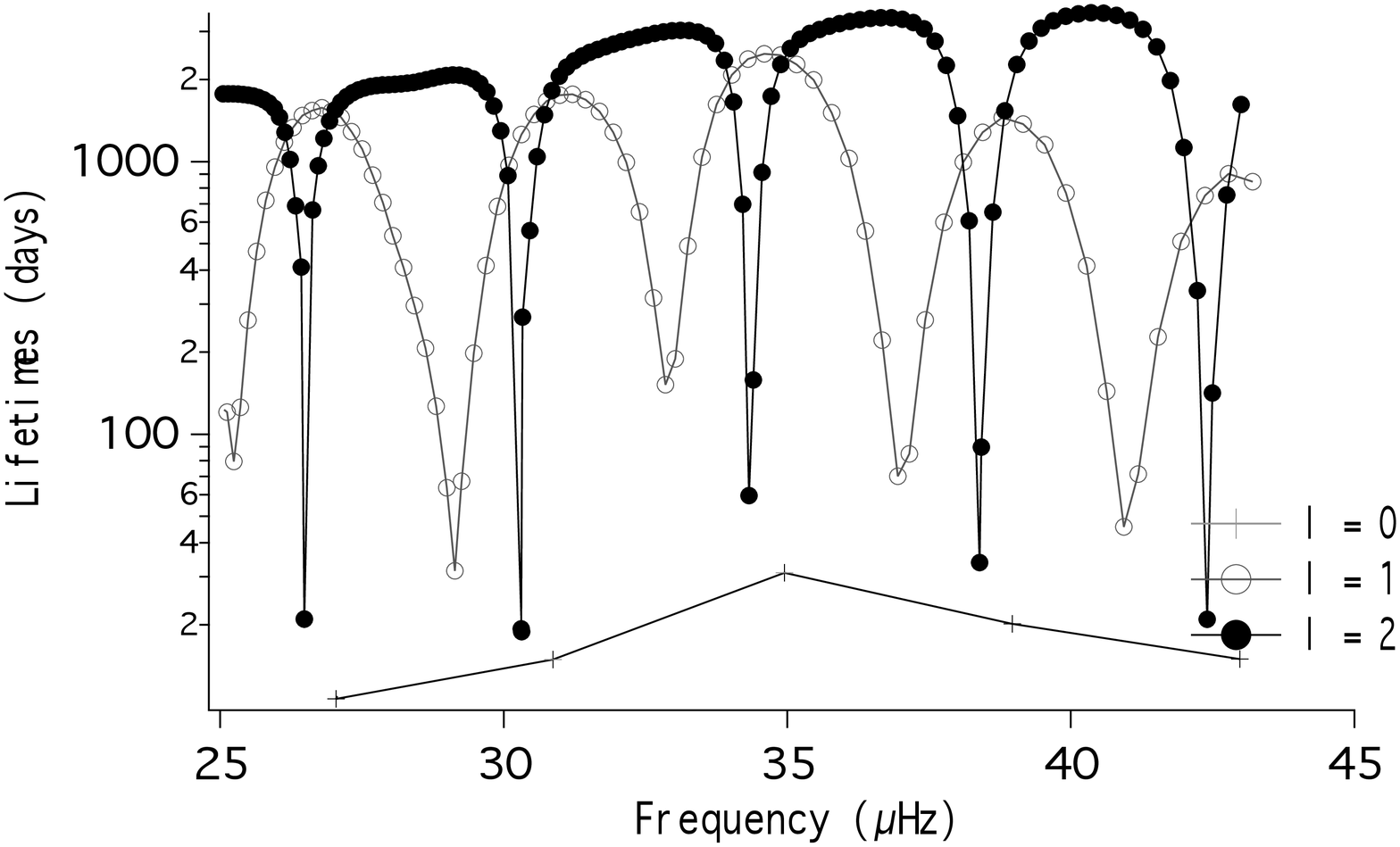} 
}
\caption{\textbf{Lifetimes} of l=0, 1 and 2 modes for the RGB model (left) and the HB model (right). 
The oscillatory behavior of the lifetime in the HB model comes from the oscillations of the inertia. It is no longer present in the RGB model due to a high radiative damping for all modes not trapped in the envelope. 
}
\label{fig:2}       
\end{figure}

\begin{figure}[h!]
\resizebox{0.9\columnwidth}{!}{
\includegraphics{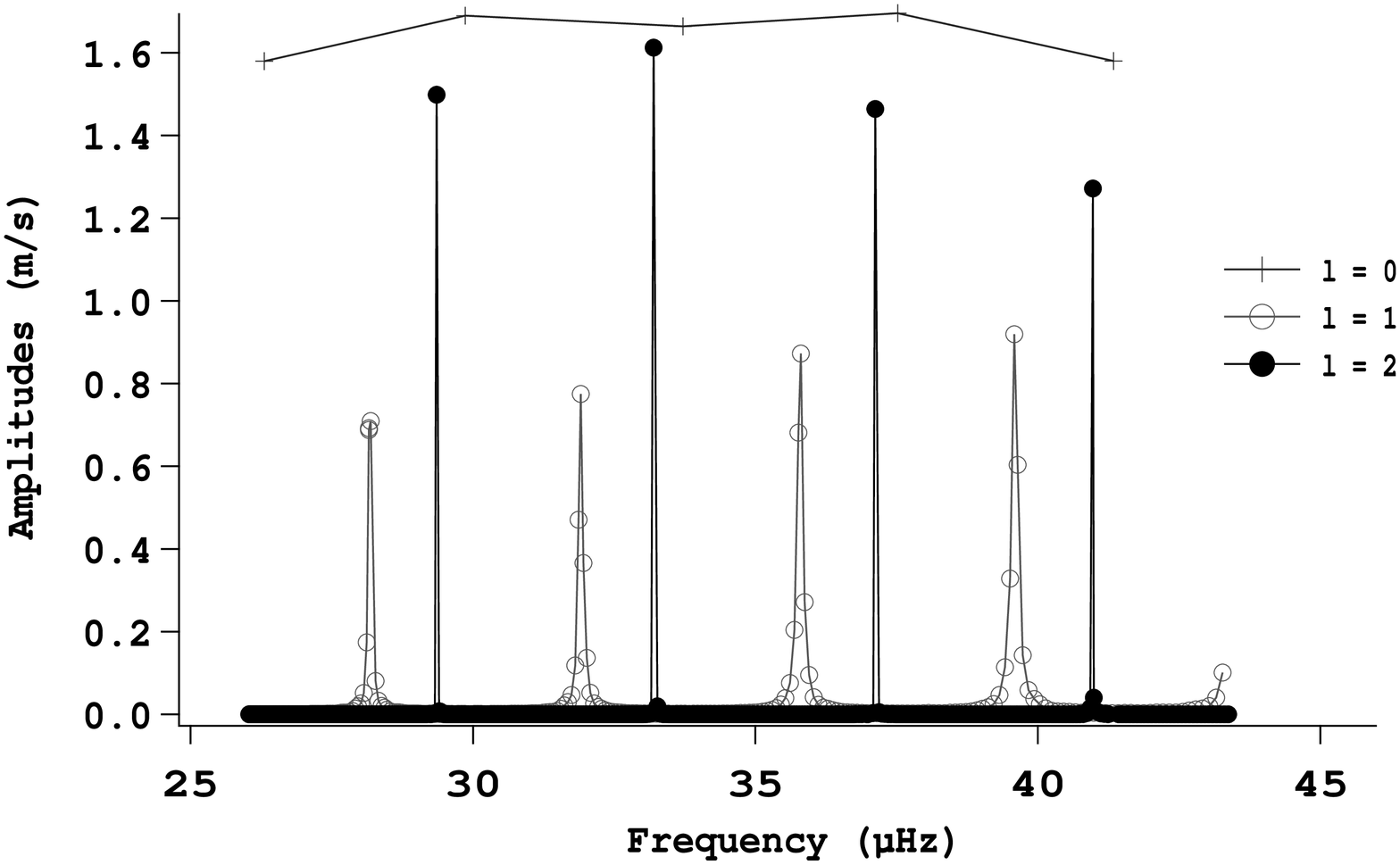} 
\includegraphics{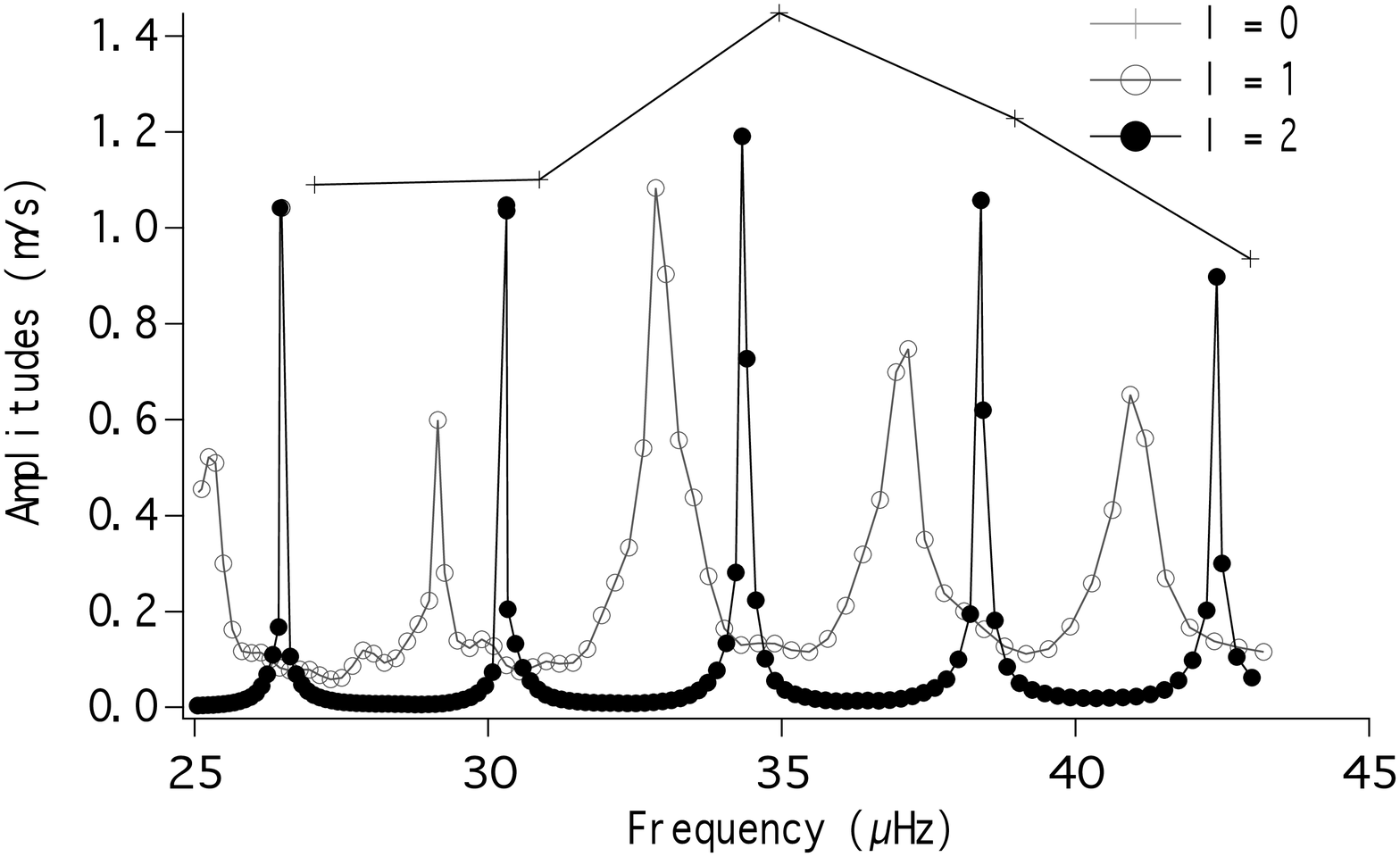} 
}
\caption{\textbf{Amplitudes} of l=0, 1 and 2 modes for the RGB (left) and the HB (right) models.
Because of larger radiative damping (left panels) and larger inertia (right panels) of l=1 modes trapped in the envelope, their amplitudes are smaller than the corresponding l=2 and l=0 modes.
The l=1 to l=0 amplitudes ratios are smaller for the RGB due to the important radiative damping. This results is not incompatible with current observations \cite{RefMosser}. 
}
\label{fig:3}       
\end{figure}

\section{Conclusions}
As already found by adiabatic computations \cite{RefMontalban}, the spectrum density is larger in the RGB model. 
Lifetimes and amplitudes indicate that more g-dominated mixed-modes should more be visible in the HB than in the RGB model.  This is due to a better trapping efficiency in the RGB model (see \cite{RefMontalban}) which results from its larger density contrast. 
Computations of lifetimes show the importance of taking into account the radiative damping to predict visibilities of mixed-modes. 
Further development and comparisons with observations will allow us to better characterise convection and its interaction with oscillations in the outermost layers of red giants.

\end{document}